\documentclass[article,
amsmath, amssymb, aps, prb,
floatfix,
twocolumn, 10pt
]{revtex4-1}

\usepackage{CJKutf8}
\usepackage{graphicx}
\usepackage{bm}
\usepackage{wasysym}
\usepackage{bbold}
\usepackage{xcolor}
\usepackage{hyperref}

\begin{document}

\title{Origins of electromagnetic anisotropy in monolayer black phosphorus}

\author{Pengke Li (\begin{CJK*}{UTF8}{gbsn}李鹏科\end{CJK*})}
\email{pengke@umd.edu}

\affiliation{Department of Physics,
University of Maryland, College Park, MD 20742}

\begin{abstract}

Contrary to empirical observations, lowest-order $\bm{k}\cdot\hat{\bm{p}}$ theory predicts that monolayer black phosphorus (``phosphorene'') is completely immune to zigzag-polarized optical excitation at the bandgap energy. Using symmetry arguments, we derive a $2\times2$ Hamiltonian under the $\bm{k}\cdot\hat{\bm{p}}$ formalism including  higher-order corrections, which is used to show that the experimentally-measured band-gap transition with zigzag polarization is dominated by the third order $\bm{k}\cdot\hat{\bm{p}}$ perturbation in the interband optical matrix element, whereas the effects of spin-orbit interaction are negligible in this material, consistent with a trivial orbital diamagnetic contribution to the  $g$-factor.


\end{abstract}

\maketitle  
\section{introduction}


Monolayer or few-layer black phosphorus is a 2-dimensional van der Waals layered material with many strong in-plane anisotropic properties.\cite{Ye_ACS2014, Li_nnano2014}
The anisotropy is endowed by the material's orthorhombic lattice structure, which can be roughly envisioned as a distorted honeycomb lattice with all the bonds parallel to the armchair direction tilted out-of-plane alternatively by approximately $\pm 72^{\circ}$, such that the otherwise-flat atomic plane is ``puckered'', with lattice constants $a_x = 4.376$~\AA~ and $a_y = 3.314$~\AA.\cite{Takao_JPSP1981} Consequently, the electronic structure is drastically different for band dispersion along the zigzag and armchair directions, reflecting the underlying geometric symmetry of the system.

Group theory was first used to investigate the band structure symmetry of monolayer black phosphorus,\cite{Pengke_PRB2014} and was later applied to incorporate external effects, such as strain\cite{LYV_NJP2015}, lattice vibration\cite{Jorio_PRB2015} and extrinsic spin-orbit coupling.\cite{Farzaneh_arXiv2019} 
In Ref.~[\onlinecite{Pengke_PRB2014}], the ``method of invariants''\cite{Lok_book} was implemented on a multiband Hamiltonian to examine lowest-order interband interactions, 
providing comprehensive understanding of the physics governing the anisotropy of features in the band structure, optical selection rules, and spin mixing. 

Monolayer black phosphorus is a semiconductor with a bandgap $\sim 2$~eV near the Brillouin-zone center ($\Gamma$-point), as given computationally by sophisticated DFT+GW calculation\cite{Tran_PRB2014,Katsnelson_GW_PRB2015,Louie_nano2017} and experimentally by photoluminescence spectroscopy. \cite{Xia_nnano2015} Since most experimentally-probed properties are closely related to the gap-edge conduction and valence states, it is most convenient to adopt a simple model focusing solely on these orbitally-nondegenerate bands, described by a $2\times2$ Hamiltonian
\begin{align}
\hat H = \begin{bmatrix}
H_\text{c} & H_\text{cv}\\
H_\text{cv}^\dagger & H_\text{v}
\end{bmatrix}.\label{eq:H}
\end{align}
Here, the basis functions are naturally chosen as the gap edge conduction and valence states at the $\Gamma$-point. It is evident from an atomic orbital tight binding model that the conduction (valence) band belongs to the irreducible representation $\Gamma_4^-$ ($\Gamma_{2}^+$) with $z$-like ($xz$-like) spatial symmetry.\cite{Takao_JPSP1981} As a result, the symmetry of the off-diagonal $H_\text{cv}$ is given by their direct product $\Gamma_2^+\otimes\Gamma_4^- = \Gamma_3^-$, which is $x$-like. Here we follow the convention in Ref.~[\onlinecite{Pengke_PRB2014}] with the $x$, $y$ and $z$ axes corresponding to the armchair, zigzag, and out-of-plane directions, respectively. 
Relevant analyses from Ref.~[\onlinecite{Pengke_PRB2014}], such as the symmetry behaviors of the $\Gamma$-point irreducible representations and their direct product relations, are organized in the Supplemental Material (SM).\cite{SM}

The $x$-like off-diagonal $H_\text{cv}$ indicates that the 
gap edge states are directly coupled by the momentum operator $\hat{p}_x$, whereas $\hat{p}_y$ is forbidden by symmetry.
This fact can also be seen from a reduced tight binding model under continuum approximation, \cite{Katsnelson_TB_PRB2015} and is consistent with the experimentally-observed strong photoluminescence with polarization along the armchair direction.\cite{Xia_nnano2015}
Surprisingly, the same experiment also demonstrated that the zigzag-polarized photoluminescence is small but certainly nonzero, which is not explained by any lowest-order model.
Recently, it was proposed that symmetry-allowed $\bm{k}$-dependent spin orbit coupling is responsible for the optical transition polarized along $y$,\cite{Fabian_arXiv2019} but the argument relies on an unreasonably overestimated spin-orbit parameter. On the contrary, spin-free \textit{ab initio} calculations show that the optical spectrum of zigzag polarization is indeed finite and increases beyond the bandgap, with amplitude two orders of magnitude smaller than that of the armchair polarization,\cite{Tran_2DM2015,Louie_nano2017} suggesting the origin of this puzzling observation should be related to higher-order contributions from the $\bm{k}\cdot\hat{\bm{p}}$ interaction.

Earlier attempts to introduce higher order $\bm{k}\cdot\hat{\bm{p}}$ components in the off-diagonal matrix element were made, but only to fit the ultraflat zigzag valence band dispersion.\cite{Neto_PRL2014} There, quadratic forms like $k_{x,y}^2$ (belonging to the scalar $\Gamma_1^+$ representation) were arbitrarily added into $H_\text{cv}$. 
Although their parameters could be chosen as complex values to maintain the necessary time reversal invariance of the Hamiltonian,\cite{Low_arXiv2017} one cannot ignore the more stringent fact that the underlying \textit{spatial symmetry} forbids the coexistence of linear and quadratic terms. 
Hamiltonians constructed with incompatible symmetry may still produce a recognizable eigenspectrum, yet the optical matrix elements derived from it will be seriously flawed. 

Note that the $\bm{k}\cdot\hat{\bm{p}}$ Hamiltonian under this symmetry argument should not be confused with the different Hamiltonian constructed from the  tight-binding approach.\cite{Katsnelson_TB_PRB2015}
In the latter case, under the continuum approximation, both linear and quadratic terms are allowed in the off-diagonal matrix element which describes the hopping between atomic sites, rather than perturbative interband coupling as in $\bm{k}\cdot\hat{\bm{p}}$. It is straightforward to verify that a unitary transform of the tight binding Hamiltonian into the band basis at the zone center eliminates any off-diagonal quadratic terms.

The present study aims to resolve any confusion surrounding the aforementioned problem. We start by deriving concise matrix elements of $\hat H$ in Eq.~(\ref{eq:H}) that capture necessary symmetry-allowed $\bm{k}\cdot\hat{\bm{p}}$ terms up to the third order in perturbation theory, including spin-orbit coupling. From the resulting interband optical matrix element, together with the band dispersion, we calculate the polarization-dependent dipole interaction strength as a function of the photon energy. 
The underlying physics of the optical anisotropy across the bandgap, especially of the weak but finite optical transition with linear polarization along the zigzag orientation, is shown to be dominated by the third-order $\bm{k}\cdot\hat{\bm{p}}$ interband matrix element, whereas the influence of spin-orbit coupling is negligible. 
At the end of the paper, the interband optical matrix elements are also used to investigate the small diamagnetic correction of the  $g$-factor induced by orbital magnetic moment from the single-particle bandstructure.

\section{Interband coupling}
We consider symmetry-allowed terms in $H_\text{cv}$ up to third order using the method of invariants,\cite{Lok_book} including invariant components of $k_x$, $k_xk_y^2$ and $\sigma_z k_y$ (the other $x$-like third-order term $k_x^3$ is not considered, as explained at the end of this section).\cite{SM} The coupling between the conduction and valence bands is then constructed as
\begin{align}
H_\text{cv} &= iPk_x + iP_3 k_xk_y^2 + \alpha \sigma_zk_y,\label{eq:Hcv}
\end{align}
with three parameters $P = 4.6$~eV$\cdot$\AA, $P_3 = -16$~eV$\cdot$\AA$^3$, and $\alpha = -5.0$~meV$\cdot$\AA, calculated from \textit{ab initio} wavefunctions using the {\sc Quantum ESPRESSO} package.\cite{QE-2009}
Note that polynomial fits to the \textit{ab initio} band dispersion cannot be used to reliably extract these parameters, especially for the higher-order $P_3$ and the relatively small $\alpha$, whose contributions to the dispersion relation are negligible (see next section). 

Two different schemes can be implemented to calculate these parameters, strictly following the definition of the $\bm{k}\cdot\hat{\bm{p}}$ formalism.\cite{Lok_book}
The values given here are obtained by projecting band edge wavefunctions at small but nonzero $\bm{k}$ onto those at the $\Gamma$ point, and the resulting off-diagonal term multiplied by $E_\text{g}$ equals the corresponding components in $H_{\text{cv}}$. In a different approach, we take advantage of the planewave basis of the wavefunctions and directly evaluate the matrix element of the momentum operators between the states at the $\Gamma$-point, resulting in parameters $\sim10\%$ larger than those given by the first method. 
We have also verified that these values are robust against variation in the details of \textit{ab initio} calculation inputs (such as the types of density functional and pseudopotential, variation in lattice constants, etc).
Details of the \textit{ab initio} process used, as well as the calculation of these three parameters, are included in SM.\cite{SM}. 
In the following we elaborate the origins of the three terms in Eq.~(\ref{eq:Hcv}).

The first term $iPk_x$ arises from the lowest order $\bm{k}\cdot\hat{\bm{p}}$ perturbation, 
\begin{align}
\hat H_{\bm{k}\cdot\hat{\bm{p}}} &= \frac{\hbar}{m_0}\bm{k}\cdot \hat{\bm{p}} =\frac{\hbar}{m_0}(k_x\hat p_x+k_y\hat p_y) .\label{eq:kp}
\end{align}
Since only the $x$ component is symmetry-allowed, we have $P= \frac{\hbar}{im_0}\langle\Gamma_{4\text{c}}^-|\hat{p}_x|\Gamma_{2\text{v}}^+\rangle$, as shown in Fig.1(a). Similar to most other semiconductors, $P$ is close to the nearly-free electron value $\pi\hbar^2/(m_0a_x)$.\cite{Cardona_book}

\begin{figure}
\includegraphics[scale=.85]{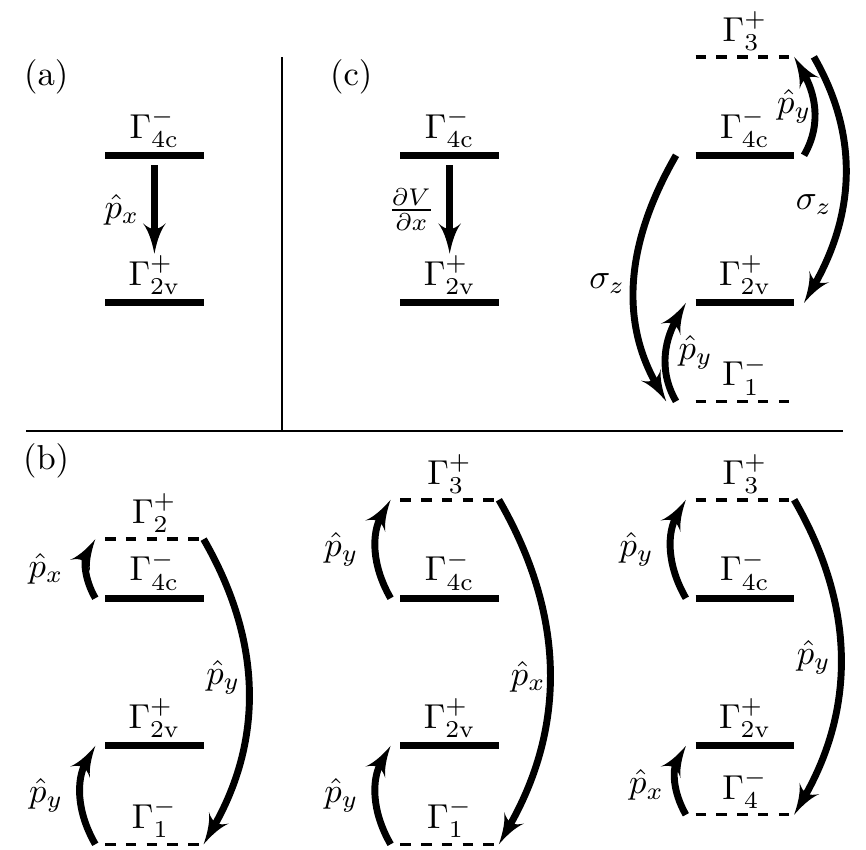}
\caption{Symmetry-allowed perturbation paths of the off-diagonal matrix element $H_\text{cv}$ between the gap edge conduction ($\Gamma_\text{4c}^-$, $z$-like) and valence ($\Gamma_\text{2v}^+$, $xz$-like) states. (a) Lowest order $\bm{k}\cdot\hat{\bm{p}}$ term [Eq.~(\ref{eq:kp})]. (b) Three types of third order $\bm{k}\cdot\hat{\bm{p}}$ paths [see Eq.~(\ref{eq:kp3}) for the left panel]. Notice the order of the $\hat{p}_x$ and $\hat{p}_y$ operators. Each of the dashed lines represents not a specific band but rather all intermediate states at remote energy sharing the same symmetry, and could be upper conduction bands or lower valence bands. (c) Left panel: $k$-dependent spin-orbit coupling [Eq.~(\ref{eq:sok})]; right panel: second order couplings via $\bm{k}\cdot\hat{\bm{p}}$ and $k$-independent spin-orbit term [Eq.~(\ref{eq:so})].
All paths share the same $x$-like spatial symmetry.  \label{fig:paths}}
\end{figure}

The second term $iP_3 k_xk_y^2$ originates from the third order $\bm{k}\cdot \hat{\bm{p}}$ perturbation via three types of pathways as shown in Fig.~\ref{fig:paths}(b), depending on the sequence of the $\hat{p}_x$ and the two $\hat{p}_y$ operators. For example, the panel on the left hand side represents the path
\begin{align}
\frac{\hbar^3}{m_0^3}\sum_{\Gamma_2^+, \Gamma_1^-}
\frac{\langle\Gamma_{4\text{c}}^-|\hat{p}_x|\Gamma_{2}^+\rangle\langle\Gamma_{2}^+|\hat{p}_y|\Gamma_{1}^-\rangle\langle\Gamma_{1}^-|\hat{p}_y|\Gamma_{2\text{v}}^+\rangle}{[E(\Gamma_{4\text{c}}^-)-E(\Gamma_{2}^+)][E(\Gamma_{1}^-)-E(\Gamma_{2\text{v}}^+)]}k_xk_y^2,\label{eq:kp3}
\end{align}
in which the summation is over all intermediate states with $\Gamma_1^-(xyz)$ and $\Gamma_2^+$ symmetries, except the highest valence band. Similarly, the other two panels in Fig.~\ref{fig:paths}(b) represent paths involving intermediate states belonging to $\Gamma_3^+(yz)$, $\Gamma_1^-$ and $\Gamma_4^-$. 

The spin-orbit term $\alpha \sigma_z k_y$ has two comparable contributions as shown in Fig.~\ref{fig:paths}(c). First of all, there is the direct $\bm{k}$-dependent spin-orbit coupling
\begin{align}
\hat H_\text{SO}^k \!&=\! \frac{\hbar^2}{4m_0^2c^2}\nabla \hat V(\bm{r})\times\bm{k}\cdot\bm{\sigma} \label{eq:sok} \\
&=\frac{\hbar^2}{4m_0^2c^2}\left[\!\left(\!\frac{\partial \hat V}{\partial x}k_y\!-\!\frac{\partial \hat V}{\partial y}k_x\!\right)\!\sigma_z\!+\!\frac{\partial \hat V}{\partial z}\!\left(k_x\sigma_y\!-\!k_y\sigma_x\right)\!\right],\nonumber
\end{align}
in which only the first term (proportional to $\frac{\partial \hat V}{\partial x}k_y\sigma_z$) has the required $x$-like symmetry and couples the conduction and valence  bands, as shown by the left panel in Fig.~\ref{fig:paths}(c). In addition, there are second-order perturbation paths via the $y$-component of $\hat H_{\bm{k}\cdot\hat{\bm{p}}}$ ($\Gamma_2^-$, $y$-like) and the $\sigma_z$ component ($\Gamma_4^+$, $xy$-like) of the $\bm{k}$-independent spin-orbit coupling term
\begin{align}
\hat H_\text{SO} &= \frac{i\hbar}{4m_0^2c^2}\nabla \hat V(\bm{r})\times\hat{\bm{p}}\cdot\bm{\sigma}.\label{eq:so}
\end{align}
The combination of the two operators has a net symmetry $\Gamma_2^-\otimes\Gamma_4^+ = \Gamma_3^-$, and the intermediate states involved belong to $\Gamma_3^+$ or $\Gamma_1^-$.

The strength of the spin-orbit term can be estimated from the ratio between the $k$-linear operators $\hat H_\text{SO}^k$ and $\hat H_{\bm{k}\cdot\hat{\bm{p}}}$, which is around $V/4m_0c^2$, with $V$ on the order of 1~keV for phosphorus core electron levels (in the region where the potential varies most drastically\cite{Chelikowsky_PRB1976}), and $4m_0c^2 \approx 2$~MeV. As a result, $|\alpha|\sim 10^{-3}{P}$, consistent with the values we obtained. For comparison, the neighboring elemental material (Si) has $P\approx 9$~eV$\cdot$\AA\, and $\alpha = 8.6$~meV$\cdot$\AA\, at the conduction band minimum.\cite{Pengke_PRL2010}

Despite their higher-order nature, both $iP_3 k_xk_y^2$ and $\alpha \sigma_zk_y$ terms in the off-diagonal Hamiltonian matrix element lead to a nonzero interband optical transition with polarization along $y$ (zigzag), since their relevant matrix elements of the momentum operator $\frac{m_0}{\hbar}\nabla_{\bm{k}}\hat{H}$ are nonzero. However, direct comparison with the experimental spectra requires integration of $\partial H_\text{cv}/\partial k_y$ over $\bm{k}$-points with the same transition energy in the Brillouin zone. Because of the ultra-flat valence band in phosphorene, the third-order term $P_3 k_xk_y^2$ dominates over the linear spin-orbit term $\alpha \sigma_zk_y$, as we will thoroughly discuss in Sec.~\ref{sec:dipole}. 

The same consideration justifies our neglect of the symmetry-alowed term $k_x^3$,  as mentioned at the beginning of this section, since $\partial k_x^3/\partial k_y$ vanishes. This third order term does contribute to the $x$-polarized dipole transition, but is negligible compared with the lowest order $k_x$-linear term.

\section{Diagonal matrix elements}
Both $H_\text{c}$ and $H_\text{v}$ have scalar symmetry ($\Gamma_1^+$). By capturing the higher-order influence from remote bands due to L{\"o}wdin folding,\cite{Lowdin_JCP1951} we have
\begin{align}
H_\text{c} &= E_\text{g}+A_\text{c} k_x^2 + B_\text{c} k_y^2,\label{eq:Hc}\\
H_\text{v} &= A_\text{v} k_x^2 + B_\text{v} k_y^2\left[\frac{C_\text{v}}{1+ (k_y/k_t)^2}-C_\text{v}+1\right].\label{eq:Hv}
\end{align}
The underlying physics of the coefficient parameters and the unusual form of $H_\text{v}$ beyond quadratic order are explained in the following. First of all, the scalar symmetry is evident in Eqs.~(\ref{eq:Hc}) and (\ref{eq:Hv}) due to the even powers of $k_x$ and $k_y$. Here, due to the well-known issue of band gap underestimation in the DFT process, the value of the $\Gamma$-point band gap used is $E_g = 0.7$~eV,\cite{Ye_ACS2014,Neto_PRL2014} which is smaller than the $\sim 2$~eV given by more sophisticated DFT+GW calculation.\cite{Tran_PRB2014,Tran_2DM2015,Louie_nano2017,Katsnelson_GW_PRB2015}  However, this quantitative issue doesn't affect the fundamental physics we focus on in this study.\cite{SM}

\begin{figure}
\includegraphics[scale=.4]{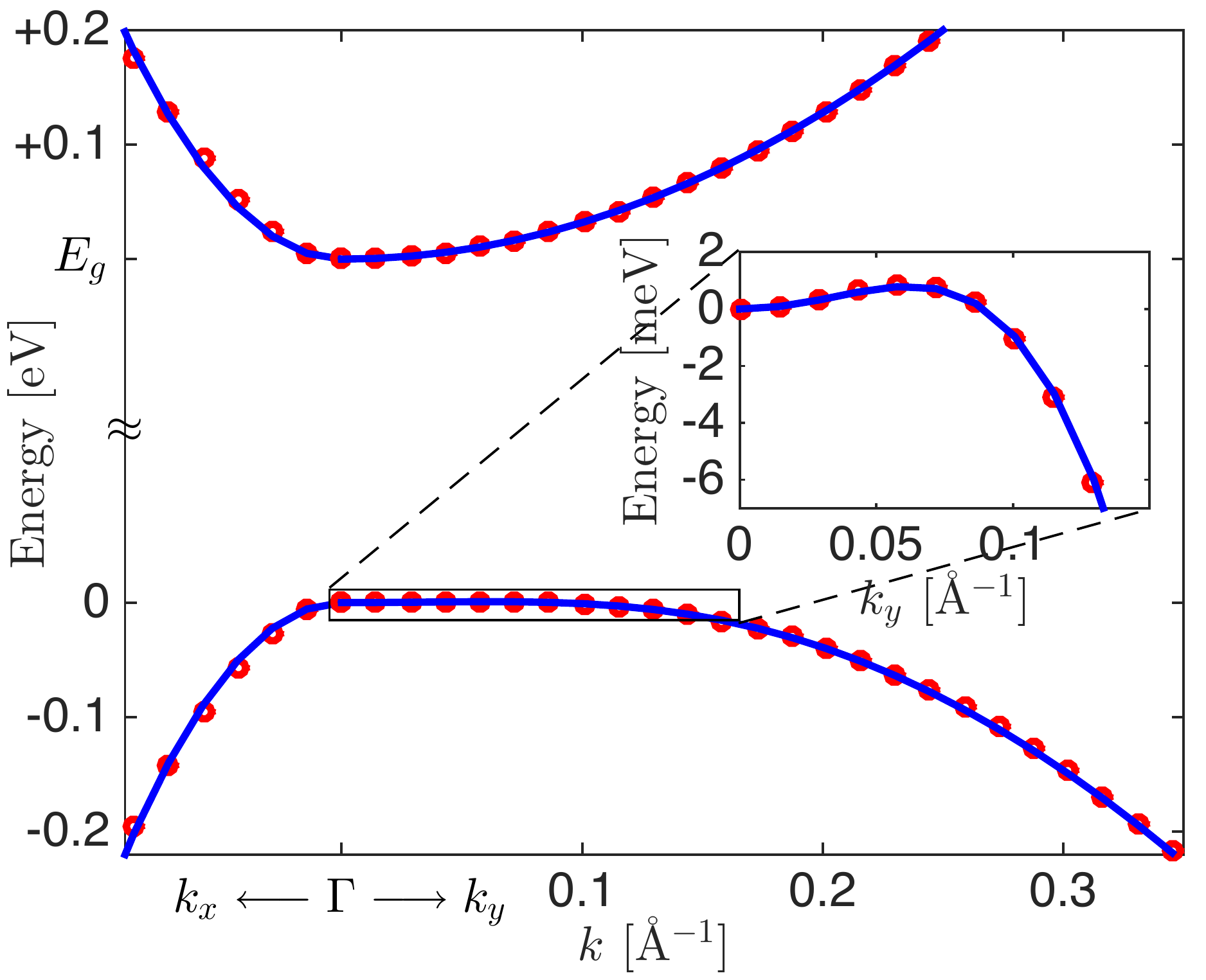}
\caption{Band structure close to the gap, along the $k_x$ (armchair) and $k_y$ (zigzag) directions. The blue curves are the spectra of the $2\times 2$ Hamiltonian detailed by Eqs.~(\ref{eq:H}), (\ref{eq:Hcv}), (\ref{eq:Hc}) and (\ref{eq:Hv}). Red circles mark the \textit{ab initio} spectra. The forbidden gap is truncated for better presentation of the features of the bands. Inset: zoom in of the valence band at small $k_y$, highlighting the unusual `electron'-like dispersion closest to the $\Gamma$-point.
\label{fig:band_fit}}
\end{figure}

Along the armchair ($k_x$) direction, $A_{\text{c,v}}$ quantify the influence from remote bands on top of the free electron dispersion,\cite{Lok_book} and together with the $iPk_x$ component in $H_\text{cv}$, determine the conduction and valence band curvatures at the zone center. With our \textit{ab initio} results of the effective masses along $k_x$ being $m_x^\text{c} = 0.148m_0$ and $m_x^\text{v} = -0.132m_0$, the dispersion coefficients $A_{\text{c,v}}$ are calculated according to
\begin{align}
A_{\text{c}}+\frac{P^2}{E_\text{g}}= \frac{\hbar^2}{2m_x^\text{c}},\text{ and }
A_{\text{v}}-\frac{P^2}{E_\text{g}}= \frac{\hbar^2}{2m_x^\text{v}},
\end{align}
giving $A_\text{c}\approx-5.2$~eV$\cdot$\AA$^2$ and $A_\text{v}\approx2.5$~eV$\cdot$\AA$^2$.
For comparison, the off-diagonal $\bm{k}\cdot \hat{\bm{p}}$ coupling contributes a dispersion coefficient $\frac{P^2}{E_\text{g}}\approx 29.6$~eV$\cdot$\AA$^2$, which is larger than the amplitudes of $A_{\text{c,v}}$, suggesting that the interband $\bm{k}\cdot\hat{\bm{p}}$ coupling directly between conduction and valence bands is primarily responsible for the relatively small effective masses in this direction.

On the other hand, the band curvatures along $k_y$ (zigzag) are dominated by the $k_y$-related terms of the diagonal elements, whereas the off-diagonal interband coupling plays a minor role. With an \textit{ab initio} value of the conduction band effective mass $m_y^\text{c}\approx1.18m_0$  we get $B_\text{c}= \frac{\hbar^2}{2m_{y}^\text{c}}-\frac{\alpha^2}{E_\text{g}}\approx 3.2~$eV$\cdot$\AA$^2$. For comparison, the contribution of the spin-orbit interband coupling to the band dispersion in this direction is $\frac{\alpha^2}{E_\text{g}}\approx 3.6\times 10^{-5}~$eV$\cdot$\AA$^2$, orders of magnitude smaller. This further verifies our previous argument that the value of $\alpha$ cannot be reliably extracted by functional fitting of the band structure.

The valence band dispersion along $k_y$ (zigzag) given by the second term in Eq.~(\ref{eq:Hv}) is distinct from conventional quadratic dispersion, in that the band is  `electron'-like with slightly positive curvature close to the origin, but eventually bends downward at larger $k_y$ (see inset in Fig.~\ref{fig:band_fit}), as observed in other \textit{ab initio} calculations.\cite{Neto_PRL2014,Ziletti_PRB2015,Tran_PRB2014}
This behavior is due to the gradually diminishing repulsion from a $\Gamma_1^-$ lower valence band as $k_y$ increases,\cite{Pengke_PRB2014} resulting in the `camel back'-shaped dispersion relation and the large average effective mass in this direction. 
Our expression of $H_\text{v}$ is simplified from Eqs. (17) and (19) in Ref. [\onlinecite{Pengke_PRB2014}]  that summarize this unusual behavior of band dispersion, as seen from the denominator $1+ (k_y/k_t)^2$ of the first term in the parenthesis in Eq.~(\ref{eq:Hv}). 
By fitting the \textit{ab initio} valence band dispersion along $k_y$, we get $B_\text{v}=0.43~$eV$\cdot$\AA$^2$, the unitless $C_\text{v} = 7.6$, and $k_t = 0.23$\AA$^{-1}$ quantifies the extension of the `electron-like' behavior along $k_y$. 
At small $k_y$, the dispersion is approximately $B_\text{v}k_y^2$ with an effective electron-like mass $8.8m_0$, whereas at large $k_y$, the dispersion is approximately $B_\text{v}(1-C_\text{v})k_y^2$ with an effective hole-like mass $-1.33m_0$. Note that here the band curvature only corresponds to the concept of effective mass at a certain $\bm{k}$ point, which should not be confused with the experimentally-determined effective mass resulting from a $\bm{k}$-space averaging over participating states. 

\section{Band structure and Density of states (DOS)}

With Eqs.~(\ref{eq:Hcv}), (\ref{eq:Hc}) and (\ref{eq:Hv}), the $2\times2$ Hamiltonian Eq.~(\ref{eq:H}) can be analytically diagonalized with the spectrum shown in Fig.~\ref{fig:band_fit}. It perfectly matches the \textit{ab initio} band structure close to the gap edge, including the ultraflat feature of the valence band along $k_y$ as shown in the inset. 
The turning point of $E_\text{v}(k_x = 0, k_y)$ is around $k_y = 0.06$\AA$^{-1}$, $\sim7\%$ from $\Gamma$ to the zone edge, indicating that states with large $k_y$ might play important roles in transport or optical phenomena. It is easy to verify that, within the energy range of our interest, the third order term $iP_3 k_x k_y^2$ has only a minor contribution to the dispersion relation, and that of the $\bm{k}$-dependent spin-orbit term $\alpha k_y$ is negligible.

\begin{figure}
\includegraphics[scale=.4]{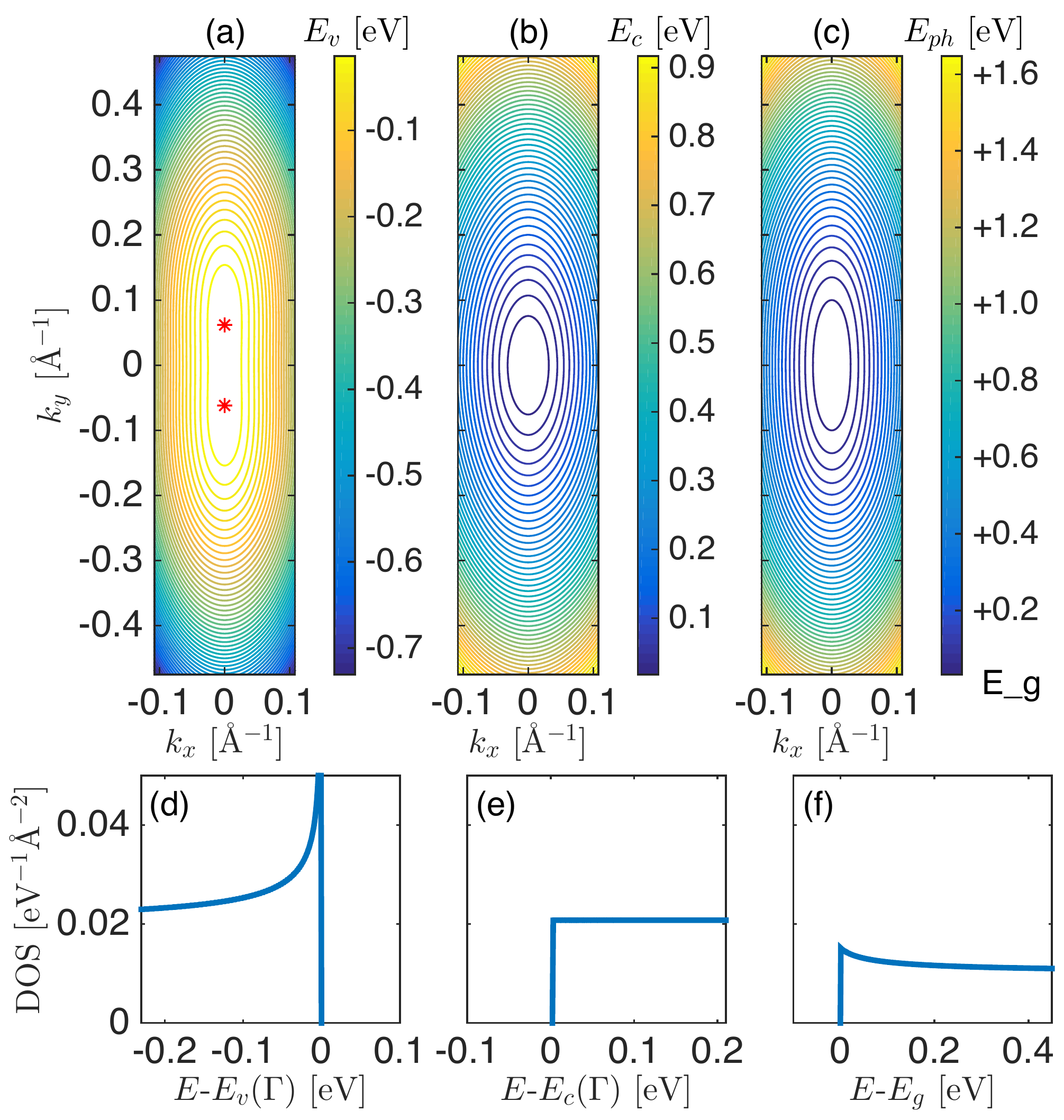}
\caption{
$\bm{k}$-space contours of the dispersion relation of (a) the valence band $E_\text{v}$, (b) the conduction band $E_\text{c}$, and (c) the optical energy $E_\text{c}-E_\text{v}$. The two absolute maxima of $E_\text{v}$ are marked by asterisks. (d) and (e) are densities of states of the valence and conduction bands, respectively, with energies measured from the band edges at the $\Gamma$ points. Note the singularity in (d) due to 
nonparabolic 
dispersion. (f) is the joint density of states of the conduction and valence bands, with energy measured from the band gap at the $\Gamma$ points.   \label{fig:DOS}}
\end{figure}

The constant-energy surface contours of the valence and conduction bands, and their energy difference across the bandgap are given in Fig.~\ref{fig:DOS}(a)-(c), respectively. The valence band contour is in an oval stadium shape with a stronger anisotropy than the elliptical shape of the conduction band isoenergetic contours. The valence band absolute maxima are not at the zone center but rather at two points along the $k_y$ axis, reproducing the dispersion shown in the inset of Fig.~\ref{fig:band_fit}.

The densities of states (DOS) of the two bands as functions of the energy can be calculated from the 2D dispersion relation according to 
\begin{align}
\text{DOS}_{\text{c,v}}(E)&= 2\int d^2\bm{k} \,\delta[E-E_{\text{c,v}}(\bm{k})],
\end{align}
where the prefactor 2 accounts for spin degeneracy. Similarly, the joint density of states (JDOS) is calculated from
\begin{align}
\text{JDOS}(E)&= 2\int d^2\bm{k} \,\delta[E-E_{\text{c}}(\bm{k})+E_{\text{v}}(\bm{k})].
\end{align}
The DOS and JDOS are presented in Fig.~\ref{fig:DOS}(d)-(f). The DOS of the conduction band is a constant around 0.021~eV$^{-1}$\AA$^{-2}$. Due to the smaller orientation-averaged effective mass, it is lower than the 2D free electron DOS of $\frac{m_0}{\pi\hbar^2} =$~0.042~eV$^{-1}$\AA$^{-2}$. In contrast, the DOS of the valence band saturates at lower energy, but approaches divergence near the band edge. Since the DOS is inversely proportional to $|\nabla_{\bm{k}} E_\text{v}|$,\cite{Haug_book} this `singularity'-like feature is the result of the unusually flat $k_y$ dispersion at the band edge, and is inherited by the JDOS close to the band gap energy.

\section{Momentum matrix element and dipole strength}\label{sec:dipole}

The interband optical matrix element could be calculated from the $2\times 2$ Hamiltonian as $ \langle \Gamma_\text{4c}^-|\frac{m_0}{\hbar}\nabla_{\bm{k}} \hat H|\Gamma_\text{2v}^+\rangle$. For simplicity, we take out the constant factor $\frac{m_0}{\hbar}$ and examine the operator $\hat{\pi}_{x,y}=\partial \hat{H}/\partial k_{x,y}$.
It is straightforward to see that $\langle \Gamma_\text{4c}^-|\hat{\pi}_x|\Gamma_\text{2v}^+\rangle= i(P+P_3k_y^2)$ and $\langle \Gamma_\text{4c}^-|\hat{\pi}_y|\Gamma_\text{2v}^+\rangle= \alpha\sigma_z+2iP_3k_xk_y$, both including a constant term and a second-order term. 
Here, we examine the contributions of each of these four terms to the optical dipole strengths as functions of the photon energy $E_{\text{ph}}$, by integration over $\bm{k}$ points with the same energy difference between the conduction and valence bands. For the two constant terms, the dipole interaction strengths are simply $P^2$ and $\alpha^2$ multiplied by the JDOS,
\begin{align}
D_{P}^x(E_\text{ph})& = 2P^2\int d^2\bm{k} \, \delta[E_\text{ph}-E_{\text{c}}(\bm{k})+E_{\text{v}}(\bm{k})], \\
D_{\alpha}^y(E_\text{ph})& = 2\alpha^2\int d^2\bm{k} \, \delta[E_\text{ph}-E_{\text{c}}(\bm{k})+E_{\text{v}}(\bm{k})],
\end{align}
where both expressions are in units of eV. Similarly, for the $\bm{k}$-dependent second order terms, we have
\begin{align}
D_{3}^x(E_\text{ph})& = 2P_3^2\int d^2\bm{k} \,k_y^4 \,\delta[E_\text{ph}-E_{\text{c}}(\bm{k})+E_{\text{v}}(\bm{k})], \\
D_{3}^y(E_\text{ph})& \!=\! 2P_3^2\!\!\int\!\! d^2\bm{k} \,4k_x^2k_y^2\, \delta[E_\text{ph}\!-\!E_{\text{c}}(\bm{k})\!+\!E_{\text{v}}(\bm{k})],
\end{align}
where the subscript `3' in $D_3^{x,y}$ indicates their origin in the third order $\bm{k}\cdot\hat{\bm{p}}$ perturbation term.
\begin{figure}
\includegraphics[scale=.32]{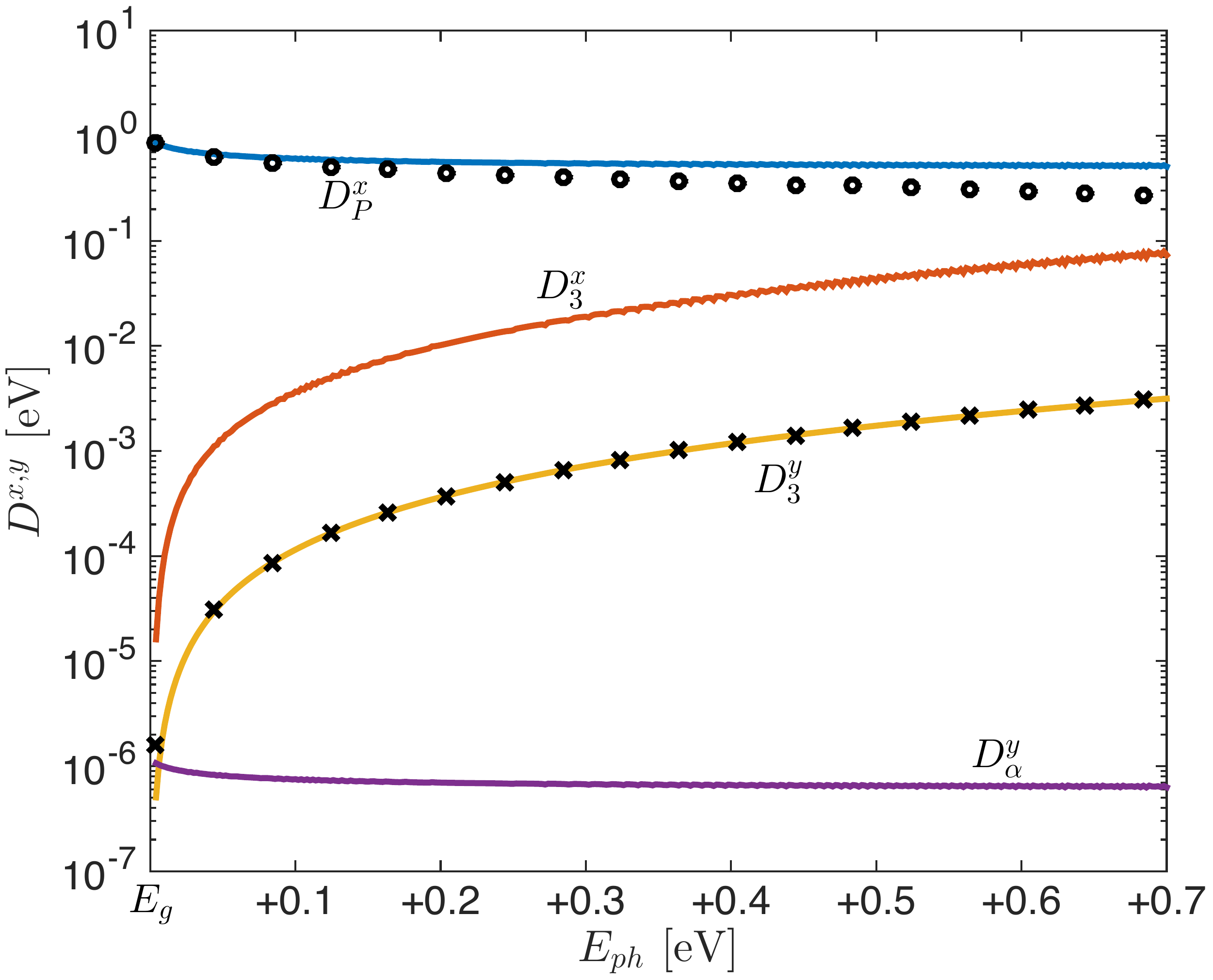}
\caption{ The four curves are interband dipole strengths given by Eqs.~(11)-(14), as functions of the photon energy beyond the bandgap. The `$\circ$' and `$\times$' markers are the total interband dipole strengths in the $x$ and $y$ directions, respectively. \label{fig:dipole}}
\end{figure}

These four quantities are plotted in Fig.~\ref{fig:dipole}, as functions of $E_\text{ph}$ beyond the band gap. As expected, $D_{P}^x$ and $D_{\alpha}^y$ are lowest order and vary only slightly throughout the energy range of our interest, in contrast with the monotonically increasing higher order $D_{3}^{x,y}$. For armchair ($x$) polarization, $D_{P}^x$ dominates over $D_{3}^x$, as expected. 
Although it is much weaker, the transition with polarization along $y$ (zigzag) is governed by $D_{3}^y$, originating from the third order $\bm{k}\cdot\hat{\bm{p}}$ perturbation paths in Fig.~\ref{fig:paths}(b). The contribution of $D_{\alpha}^y$ due to spin-orbit coupling is orders of magnitude smaller, once the photon energy is slightly beyond $E_\text{g}$. This is not only a consequence of the weak spin-orbit coupling of phosphorus, but also because interband transition at higher photon energy involves states with relatively large $k_y$, along which the dispersion is very flat. It is easier to understand this point 
from a quantitative estimation,
with the help of the energy contour in Fig.~\ref{fig:DOS}(c): for photon energy $0.7$~eV beyond the bandgap at around $\bm{k}=(0.05$\AA$^{-1},0.3$\AA$^{-1})$, we have $2P_3k_xk_y\approx0.48$~eV$\cdot$\AA$\gg|\alpha|$.

The total dipole strengths in the two orthogonal directions are given by the markers in Fig.~\ref{fig:dipole}. $D^y$ exactly equals the sum of  $D_\alpha^y$ and $D_3^y$, yet mostly overlaps with the latter, as we analyze.  $D^x$ is slightly smaller than the sum of $D_P^x$ and $D_3^x$, due to the cross product of $P$ and $P_3k_y^2$ in $|\langle\hat{\pi_x}\rangle|^2$. At high photon energy, the ratio between $D_{3}^y$ and $D_{P}^x$ is close to $\sim1\%$, which is consistent with the experimental observation of photoluminescence from decay of direct excitons,\cite{Xia_nnano2015} whose wavefunction in the 2D monolayer is spatially localized along the zigzag direction\cite{Tran_2DM2015} and involves relatively large $k_y$ components in reciprocal space.

\section{$g$-factor}
The momentum operator $\hat{\pi}_{x,y}$ could also be used to evaluate the orbital magnetic moment contribution to the Land\'{e} $g$-factor under out-of-plane magnetic field\cite{Lok_book}
\begin{align}
\Delta g\!=\! \frac{2}{im_0}\frac{m_0^2}{\hbar^2}\!\sum_{n'}\! \frac{\langle n|\hat{\pi}_x|n'\rangle\langle n'| \hat{\pi}_y|n\rangle\!-\!\langle n|\hat{\pi}_y|n'\rangle\langle n'| \hat{\pi}_x|n\rangle}{E_n-E_{n'}}.\label{eq:gfactor}
\end{align}
Within a two-band Hamiltonian, $\Delta g$ is the same for both the conduction and valence states. Letting $|n\rangle =|\Gamma_\text{4c}^-\rangle$ and $|n'\rangle =|\Gamma_\text{2v}^+\rangle$, we have 
\begin{align}
\Delta g=  \frac{4m_0}{\hbar^2}\frac{(P+P_3k_y^2)\alpha}{E_\text{g}}\approx \frac{4m_0}{\hbar^2}\frac{P\alpha}{E_\text{g}} = -0.017.\label{eq:Delta_g}
\end{align}

It is straightforward to apply Eq.~(\ref{eq:gfactor}) to energetically remote states beyond the $2\times 2$ model to calculate the complete single-particle correction to the $g$-factor. Only those states with relevant symmetries need to be taken into account. 
On one hand, the conduction (valence) state $|\Gamma_\text{4c}^-\rangle$ ($|\Gamma_\text{2v}^+\rangle$) 
can couple to any other states with $\Gamma_\text{2}^+$ ($\Gamma_\text{4}^-$) symmetry via $\hat{p_x}k_x$ and $\frac{\partial V}{\partial x}k_y$, as demonstrated by Fig.~\ref{fig:paths}(a) and the left panel of Fig.~\ref{fig:paths}(c), respectively. 
For each of these remote bands, interband coupling parameters similar to $P$ and $\alpha$, as well as the energy difference, can be calculated in the same way using the \textit{ab initio} wavefunctions, and then their contribution to the $g$-factor correction is obtained using Eq.~(\ref{eq:Delta_g}). On the other hand, as shown in the right panel of Fig.~\ref{fig:paths}(c), one should not ignore\cite{Fabian_arXiv2019} the fact that $|\Gamma_\text{4c}^-\rangle$ ($|\Gamma_\text{2v}^+\rangle$) could also couple to remote states with $\Gamma_\text{3}^+$ ($\Gamma_\text{1}^-$) symmetry, via the $\bm{k}\cdot\hat{\bm{p}}$ operator $\hat{p_y}k_y$, together with the $\bm{k}$-dependent spin orbit coupling component in Eq.~(\ref{eq:sok}) proportional to $\frac{\partial V}{\partial y}k_x$. Although these two operators are irrelevant to the gap edge coupling matrix element $H_\text{cv}$ in our $2\times 2$ model, they contribute similar $g$-factor correction as their corresponding orthogonal counterparts.

Summing over the contribution from all symmetry-related bands, the orbital magnetic moment induced total $g$ factor correction for the edge states are calculated as $\Delta g_\text{c} = -0.05$ and $\Delta g_\text{v} = -0.02$, both being diamagnetic as in regular semiconductor systems,\cite{Yafet_1963} consistent with the measured magnetic susceptibility,\cite{anie2016} and slightly reducing the $g_0=2$ of free electrons.  It is not surprising that the single-particle bandstructure-induced $g$-factor deviations are relatively small, due to 
i). the weak spin-orbit coupling in atomic phosphorus, and 
ii). orbital non-degeneracy of all the bands, as the space-group is Abelian and irreducible representations are all one-dimensional.\cite{SM} The latter is in contrast with cases in which spin-splitting of otherwise orbitally-degenerate bands could significantly reduce $g_0=2$ and even reverse its sign such as in many III-V semiconductors.\cite{Bastos_JAP2018}
Our theoretical analysis is consistent with the $g$-factor  measured recently by quantum oscillation experiment.\cite{Li_nnano2016} 
Note that in different multilayer systems under strong-correlation condition, it is suggested that electron-electron interaction could substantially modify the $g$-factor.\cite{Zhou_PRB2017}

\section*{Acknowledgement}

This study is supported by the National Science Foundation under Award No. 1707415.

\bibliography{Phosphorene2x2.bib}

\end{document}